# DEVELOPMENT OF TEST MATERIALS FOR ASSESMENT BROADCASTING VIDEO PATH

**Nameer Hashim Qasim [1)], Volodymyr Pyliavskyi [2)], Valentina Solodka[2)]**

*1) AL-Qalam University College₁ Departament of Computer Techniques Engineering nameer.qasim@icloud.com*
*2) O.S. Popov Odessa National Academy of Telecommunications₂ Department radio and television broadcasting v.pilyavskiy@ukr.net*

*The analysis of metrological agents for the estimation of the quality of telecommunication path treatments is carried out. Analysis of subjective and objective estimation methods is presented. The justification for choosing an objective method of measuring is presented. Are reported a list of existing goals and objectives, which face the current progress of the implementation of promising systems for broadcasting and improvement of existing ones. The latter include the absence of normative documents and recommendations for optical specimens of testing tables, etc. Researches related to the choice of optical samples from existing sets of colors from a number of international documents and recommendations are presented. Deficiency of these sets is substantiated, as some of the colors necessary for evaluation of telecommunication paths cannot be provided with no color of sets. Provides recommendations and refinements about which colors from existing sets should be used and what constraints exist when used and how to solve. It is shown, that saturated colors of green and red colors are not provided with spectral from the existing sets, therefore, it is proposed to expand the traditional understanding of metrological means by using atlases of colors. Satin colors In contrast to existing ones have a number of advantages and can be used for evaluation in advanced conditions, rather than in the studio that is rigidly defined by the regulations. It is noted that for different brightness values there are a different number of testing points. The number of testing points directly depends on the evaluation criteria, is driven by several possible sets of colors for evaluation. Recommendations for other evaluation conditions are provided. Use satin colors to construct testing tables*

## 1. Introduction

The rapid development of video technologies – speech, special purpose, industrial – requires all stricter conditions to the quality of the transmitted content. When a river comes about quality, it should be understood the complex of a number of subjective and objective indicators. At the moment there are quite a lot of methods of evaluating quality. The latter should be attributed to the most common-the subjective methods [ITU500], and objective methods [EBU3350, EBU3333, ITU601,]. The subjective methods of them are quite expensive, and the receipt of the result is quite extended in time. Therefore it is worth considering measuring material methods, allowing to assess in a fairly short period of time to receive results. What is relevant considering the fact that existing sets of testing material are not sufficient, and their optical spectral distribution is not defined.

## 2. Analysis of literature data and problem statement

Objective methods of quality measurement have a lot of testing material and tools for obtaining results. Their diversity is caused by a large number of devices and systems that require evaluation. One of the common measuring materials, which is the basis for subsequent metrological instruments, can be considered [1, 2]. Another work, [3] focuses on an objective evaluation of systems with an extended dynamic range and an extended area of the transmitted colors, but all are received estimates are estimates made for standard conditions without the use of variations. Currently, it is necessary to take into account different variants of measurement, including the use and optical tables, which are not marked in these works. The emergence of work [4] is aimed at evaluating the quality of ultra-high-definition 4k, 8k systems, but also by the subjective method, but with the use of images. Proposed subjective method of quality measurement is quite costly in time, and have generalized character. In the real environment, a set of images in a different plot is required for a full evaluation, which requires an increase in the measurement time with an increase in the image set. For this reason, you should use the test sets of colors included in the test image to quickly obtain estimates. Also in these works is offered only a set of nine colors, including the main and additional colors of 90% saturation. A set of a such number of colors has been proposed for a system that is not widely used at the present time, so the data presented is of questionable relevance. There is a number of Works [5 – 7], which indicate this set, but not optical colors, and artificially generated levels of signals that transmit them through communication channels. This type of signal allows evaluating the transmission tract without end-converting devices.

Therefore, optical samples should be used for evaluation of the vitreous tract, with the defined spectral distribution. Among the existing known sets of spectral color, distribution are [8]. The set presents a large number of spectra, and what needs to be used for the evaluation is unclear. One additional set may be noted [9], which is represented by significantly fewer spectral distributions, but its use is not indicated for the particular measurement case.

At the moment, there are other methods for defining the spectral distribution of the desired color [10]. This method is universal and allows to define the spectral distribution and make it a spectral sample for measuring. This procedure quite long in time and requires significant costs for the production of the sample, but it allows quite accurately set the optical parameters of the sample.

The unsolved task remains the definition which should be the spectral distributions of colors for estimation of the quality of operation of the through telecommunication pathes.

## 3. Purpose and objectives of the study

The aim of the study is to develop a test material of colorimetric evaluation of video systems functioning. This will allow you to quickly assess the functioning of video transfer systems in terms of color reproduction and construction of new systems.

To achieve the goal, the following tasks were set:
– To specify the minimum permissible set of metrological support;
– Identify possible options for evaluating colorimetric characteristics of

multimedia pathes;
 – Take into account the dynamism of the evaluation conditions.

**4. Materials and methods of study of speech estimation mean**

To achieve the tasks you should use optical samples with the known spectral distribution. Since the recommendations and unambiguous decision on the use of a definite set of spectra are not, so are proposed by solving the problem of finding such spectra. The search is performed by solving a mathematical problem of finding lower extremism of functional dependence, which expresses the distance between the coordinates reference and created a range of color. Assuming that existing spectral distributions do not fully meet the purpose, it is proposed to use the color atlas and to search for their spectral distribution to use the algorithm [10].

**4.1. Search for a minimum allowable set of colors and their spectra**

Finding the optimal colors is the problem of finding the minimum extremum of functional dependency (1).

$$\Delta E = \sqrt{\left(x_{etalon} - x_{optimal}\right)^2 + \left(y_{etalon} - y_{optimal}\right)^2 + \left(z_{etalon} - z_{optimal}\right)^2}, \quad 1$$

where $x_{etalon}$, $y_{etalon}$, $z_{etalon}$ are the color coordinates defined as reference colors. The values of the reference colors are presented in the table. 1. $x_{optimal}$, $y_{optimal}$, $z_{optimal}$ are color coordinates derived from absolute color coordinate values (2).

$$[x, y, z]_{optimal} = [X, Y, Z]_{optimal} / \sum [X, Y, Z]_{optimal}. \quad 2$$

This value $[X, Y, Z]_{optimal}$ is derived from the expression (3).

$$\{X, Y, Z\}_{optimal} = \sum_{360}^{720} S(\lambda) \cdot P(\lambda) \cdot \{\bar{x}, \bar{y}, \bar{z}\}_{CIE}. \quad 3$$

In the expression (3) The spectral distribution of the optimum optical sample $S(\lambda)$, $P(\lambda)$ describes the spectral distribution of the light source, and $\{x_{10}, y_{10}, z_{10}\}_{CIE}$ – Spectral characteristics of the camera COLOR bands according to CIE Standard.

The task of minimization was solved using the search function of such spectral distribution $S(\lambda)$ for which $\delta\varepsilon \leq 10^{-5}$.

```
% x0 reference point coordinates, y0, z0 are set in the Funerror function

delta_E = optimset (' Display ', ' off ', ' TolX ', 1e -5);
Lambda = Fminsearch (@funerror, [490.545], delta_E);

Disp ([' lambda1 = ', Num2str (lambda (1))])
Disp ([' Lambda2 = ', Num2str (Lambda (2))])
```

This algorithm is proposed to be performed in the Matlab modeling environment [11] using the following script. The script is the main program and includes auxiliary programs implemented with the use of functional dependencies (1) – (3).

**4.2. Possible options for evaluating colorimetric characteristics of multimedia pathts**

The measuring materials used for the evaluation of colorimetric evaluation should include test images. Test images consist of colors that are defined by coordinates color. But this is not enough when it comes to optical images, the colors of which should be uniquely defined in terms of spectral distribution. That is why it seems to determine optimal spectral sets, what they should be in perfect form and real spectral distributions.

**4.2. 1. Optimum colors and their spectral distribution**

The limit options for implementation of the Optical test table (OVT) can be accepted by the table whose elements correspond to the optimum colors of the first and second genus [10]. Optimal colors correspond to the equivalent spectral radiation in the Spectrum area (4)

$$S(\lambda) = \begin{cases} K & \text{при } \lambda \in \overline{\lambda_1; \lambda_2}, \\ 0 & \text{при } \lambda \notin \overline{\lambda_1; \lambda_2}, \end{cases} \qquad 4$$

For colors $Ye, C, g, g_{0.5} Ye, C, G, G_{0,5}$. In the form of two plots (5)

$$S(\lambda) = \begin{cases} K & \text{при } \lambda \in \overline{\lambda_{360}; \lambda_1} \,\&\, \lambda \in \overline{\lambda_2; \lambda_{720}}, \\ 0 & \text{при } \lambda \notin \overline{\lambda_{360}; \lambda_1} \,\&\, \lambda \notin \overline{\lambda_2; \lambda_{720}}. \end{cases} \qquad 5$$

For colors $R, R_{0.5}, b, b_{0.5}, M$, where $\lambda_{360}$ and $\lambda_{720}$ are blue and red Spectrum boundaries $\lambda_{360}$= 360 nm and $\lambda_{720}$= 720 nm. In table 1, the following parameters $\lambda_1, \lambda_2, K$, as well as the relative brightness values *of $L_C$* colors corresponding to the HDTV standard, according to which $K$ values are calculated. Wavelengths and coordinates are shown in Fig. 1.

Table 1

Optimum color Data

|   | R | G | B | Ye | C | M | $R_{0.5}$ | $G_{0.5}$ | $B_{0.5}$ | WW |
|---|---|---|---|----|----|----|-----|-----|-----|------|
| R | 1 | 0 | 0 | 0.5 | 0 | 0.5 | 0.667 | 0.167 | 0.167 | 0.333 |
| G | 0 | 1 | 0 | 0.5 | 0.5 | 0 | 0.167 | 0.667 | 0.167 | 0.333 |
| B | 0 | 0 | 1 | 0 | 0.5 | 0.5 | 0.167 | 0.167 | 0.667 | 0.333 |
| $L_C$ | 0.213 | 0.715 | 0.072 | 0.464 | 0.394 | 0.142 | 0.272 | 0.524 | 0.202 | 1.000 |
| X | 0.64 | 0.30 | 0.15 | 0.4193 | 0.2246 | 0.3209 | 0.4403 | 0.3058 | 0.2242 | 0.3127 |
| Y | 0.33 | 0.60 | 0.06 | 0.5053 | 0.3287 | 0.1542 | 0.3293 | 0.4758 | 0.1827 | 0.3290 |
| $\lambda$, NM | 611 | 547 | 464 | 569 | 491 | – | 611 | 547 | 464 | – |
| $\lambda_1$, NM | 412 | 481 | 497 | 480 | 378 | 496 | 445 | 460 | 526 | 360 |
| $\lambda_2$, nm | 584 | 592 | 660 | 609 | 591 | 585 | 545 | 608 | 612 | 720 |
| $K \times 100$ | 0.823 | 0.869 | 0.858 | 0.908 | 0.929 | 0.875 | 0.439 | 0.560 | 0.574 | 0.00946 |

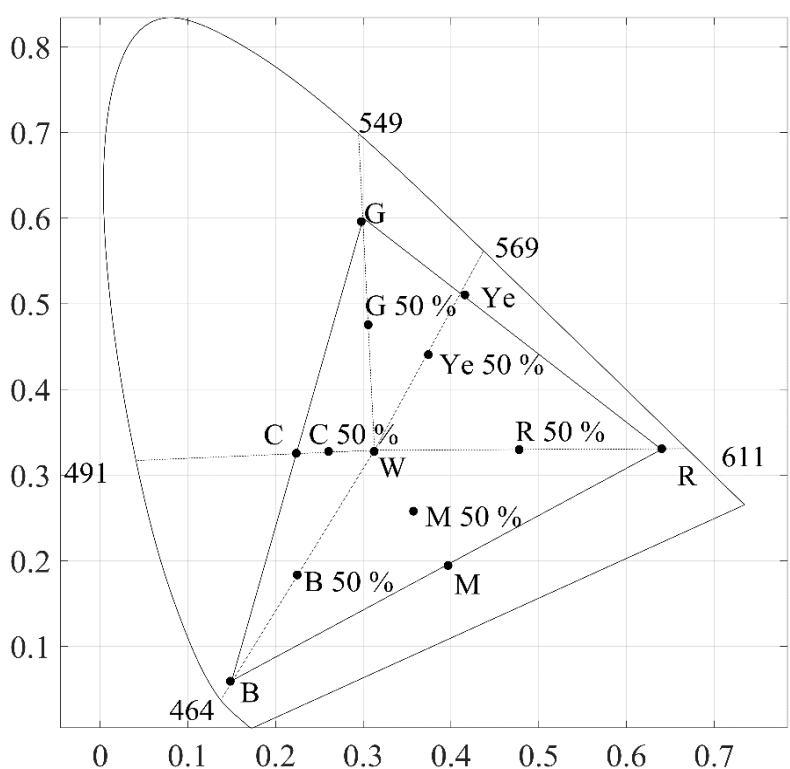

Figure. 1. Determination of central lengths of waves of Optimum colors

It should be borne in mind that in publications and documents in the field of TV technology is used the definition of relative brightness normalized to the unit interval, ie $Y \in (0...1)$. Accordingly, the brightness testing colors in the table. 1 equals $Y \in \overline{0;1} L_C$ $\in (0...1)$. In the works devoted to models of color perception, the relative brightness calculation is used normalized to the level of 100, ie $L_C \in \overline{0;1} Y \in (0...100)$. In Fig. 2 the relative brightness designation for each $Y \in \overline{0;100} L_C \zeta$ color, where $\zeta = 100$, 75 and 25 are defined as the share of greatest possible brightness for the color.

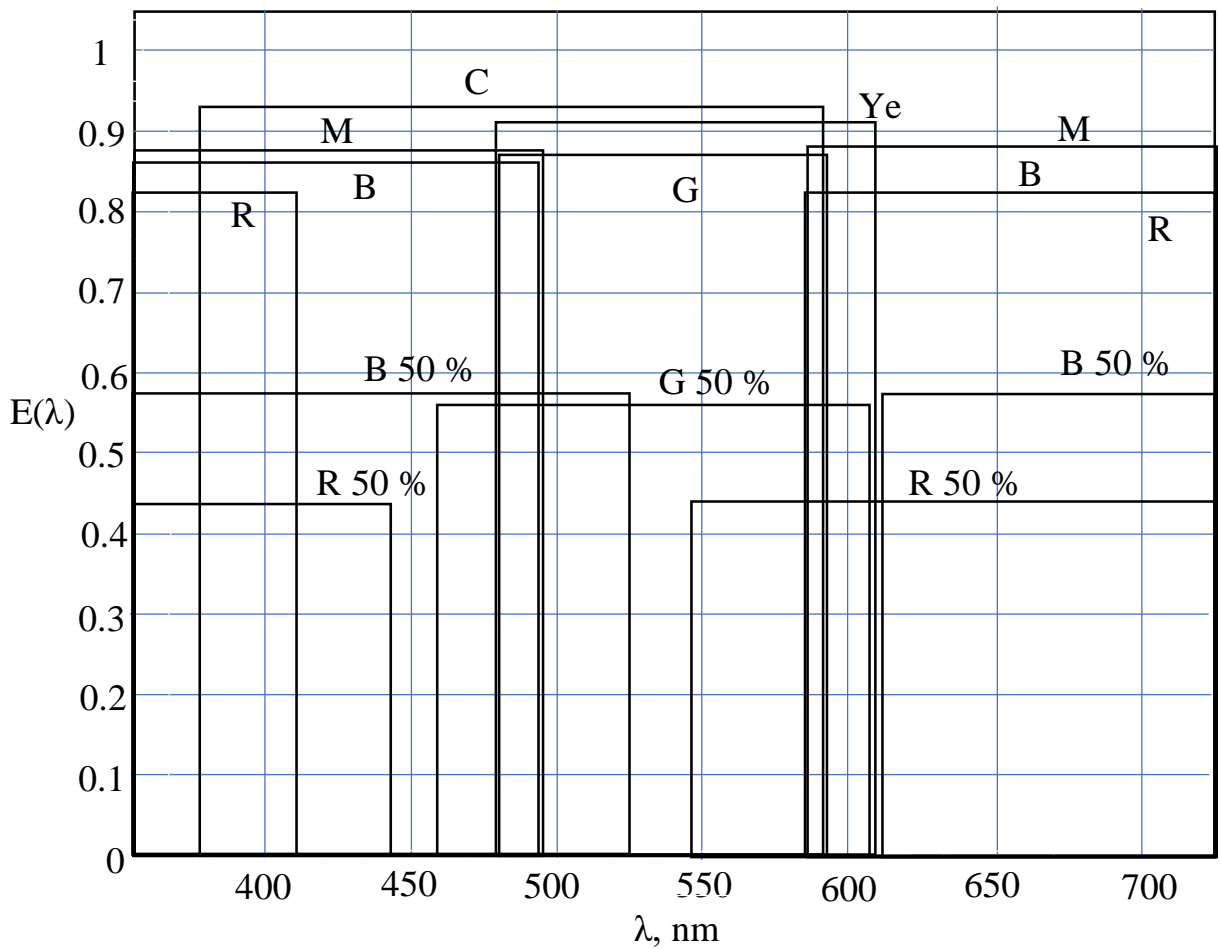

Figure. 2. Spectral distribution of optimum colors

A set of optimum spectral distributions is introduced for high-definition digital systems. Colors with a saturation of 50% are included in the color set, as they specify the colorimetric score.

But in practice, spectral distributions differ from optimal, so one of the methods of constructing spectral distributions, which is approximate to real is the study [10, 12]. It should be understood that the specified method eliminates the possibility of investigation eg METEMERIZMU phenomenon.

**4.2. 2. Variants of using real spectral color distributions**

In previous sets, the variant of using of a set of colors with the saturation of 100, 75 and 25% were considered. This set is not complete and therefore does not allow to estimate the saturated colors to the fullest. Limited is that an estimation of saturated colors cannot be made because the evaluation area is limited by the color triangle and the evaluation sector is limited by its acute angle. If you consider 75% saturation, we may note that this is not enough to estimate the saturated colors. A somewhat different approach is offered in these interpretations.

Since the implementation of optimal colors in practice is quite problematic, so these spectra are an example of what should be approached. However, for real conditions, the colors of real objects and those that are already known range are used. Therefore, the known spectral distributions of objects that are represented in the documents [8, 9, 13], to choose such colors can satisfy the set of reference colors. Spectral distributions of a set of colors of testing material for high-definition multimedia systems are proposed.

Table. 2 color markers have been marked with pairs that have been selected for both colors due to the absence of other spectra. In practice, choose a color to test the color with the smallest error.

Presented in the table. 2 results can be used to construct optical metrological support to evaluate the quality of the transmission operation of video information.

Table 2

Recommended color data for display color assessment ($x$, $y$ – High Definition TV color, $x_{spectral}$, $y_{spectral}$– Color coordinates equivalent to the most bottom spectrum)

|  | Selection of colors from the spectra set according to [8] | | | | Selection of colors from spectra set according to [9] | | | |
|---|---|---|---|---|---|---|---|---|
|  | $X_{Spectral}$ | $Y_{Spectral}$ | Color ID | $\Delta E$ | $X_{Spectral}$ | $Y_{Spectral}$ | Color ID | $\Delta E$ |
| $R$ | 0.6359 | 0.3299 | Pr_ds_2 (2) | 0.001 | 0.624 | 0.351 | 164 | 0.045 |
| $G$ | 0.3441 | 0.5040 | ph03_t (216) | 0.002 | 0.343 | 0.514 | 1492 | 0.233 |
| $B$ | 0.1742 | 0.1014 | ph03_t (228) | 0.007 | 0.199 | 0.136 | 2386 | 0.083 |
| $C$ | 0.2411 | 0.3150 | ph03_t (47) | 0.001 | 0.210 | 0.286 | 3328 | 0.024 |
| $M$ | 0.3084 | 0.1786 | Pr_ds_3 (4) | 0.006 | 0.325 | 0.171 | 3321 | 0.106 |
| $Ye$ | 0.4194 | 0.4607 | Pr_sh_2 (73) | 0.000 | 0.398 | 0.483 | 1895 | 0.045 |
| $R_{0.9}$ | 0.6359 | 0.3299 | Pr_ds_2 (6) | 0.006 | 0.607 | 0.349 | 158 | 0.011 |
| $G_{0.9}$ | 0.3441 | 0.5040 | ph03_t (216) | 0.002 | 0.343 | 0.514 | 1492 | 0.205 |
| $B_{0.9}$ | 0.1742 | 0.1014 | ph03_t (228) | 0.005 | 0.228 | 0.200 | 3099 | 0.006 |
| $C_{0.9}$ | 0.3247 | 0.1734 | Pr_ds_2 (4) | 0.001 | 0.264 | 0.332 | 2832 | 0.004 |
| $M_{0.9}$ | 0.2411 | 0.3150 | ph03_t (43) | 0.002 | 0.264 | 0.332 | 2832 | 0.097 |
| $Ye_{0.9}$ | 0.4194 | 0.4607 | Pr_sh_2 (73) | 0.002 | 0.398 | 0.483 | 1895 | 0.021 |
| $R_{0.5}$ | 0.4392 | 0.3287 | Silk (475) | 0.000 | 0.413 | 0.331 | 3527 | 0.034 |
| $G_{0.5}$ | 0.3093 | 0.4672 | gr_s (393) | 0.002 | 0.336 | 0.478 | 3436 | 0.093 |
| $B_{0.5}$ | 0.2246 | 0.1838 | Pr_sh_1 (5) | 0.001 | 0.228 | 0.200 | 3099 | 0.006 |
| $W$ | 0.3110 | 0.3322 | Pa_o (41) |  | 0.3127 | 0.3313 | 3091 | 0.003 |

In Fig. 3, 4 is shown + – colors defined for color assessment, and + – those that are found according to the reference. It is not difficult to notice the difference in the placement of coordinates, first of all, it is due to the fact that Sets have a limited number of sets of spectral distributions. Secondly, some colors are not advisable to use for evaluating saturated colors, namely green and blue areas. As for not saturated colors, the set more satisfies the presented in Fig. 3. the

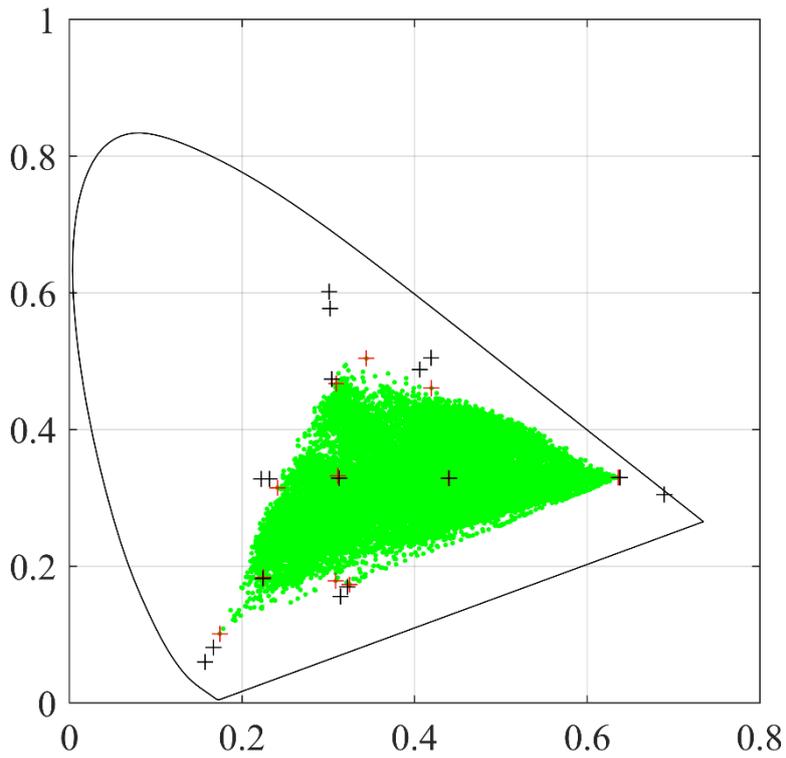

Figure. 3. Colors from the set that can be selected for evaluation

As for the set of colors represented in Fig. 4, this set provides even fewer requirements for a set of testing colors. But still, they can be used to replace those spectral distribution of which is unknown.

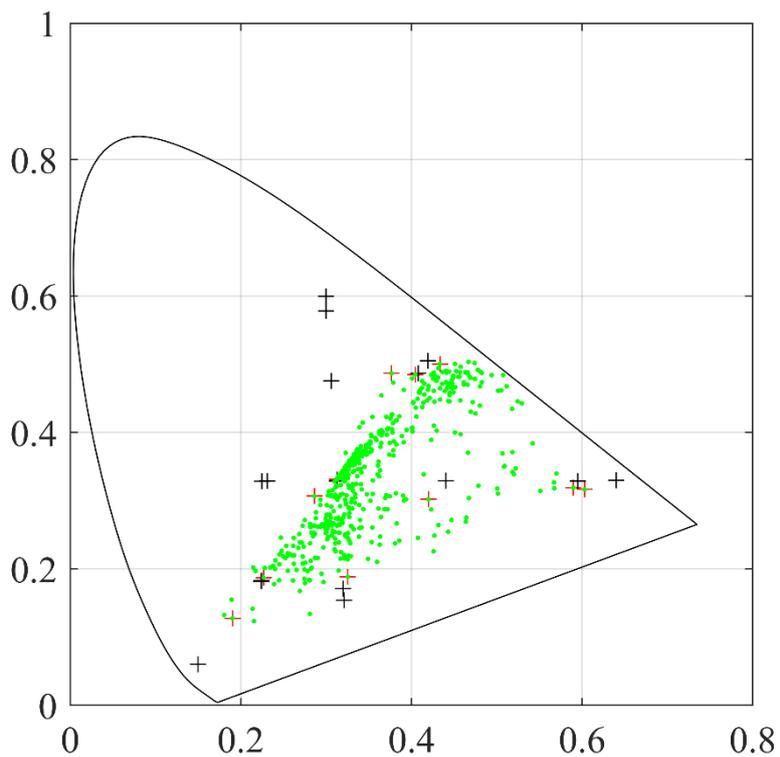

Figure. 4. Colors from the set that may be selected for evaluation

Spectral distributions of represented color points are represented in Fig. 5, 6.

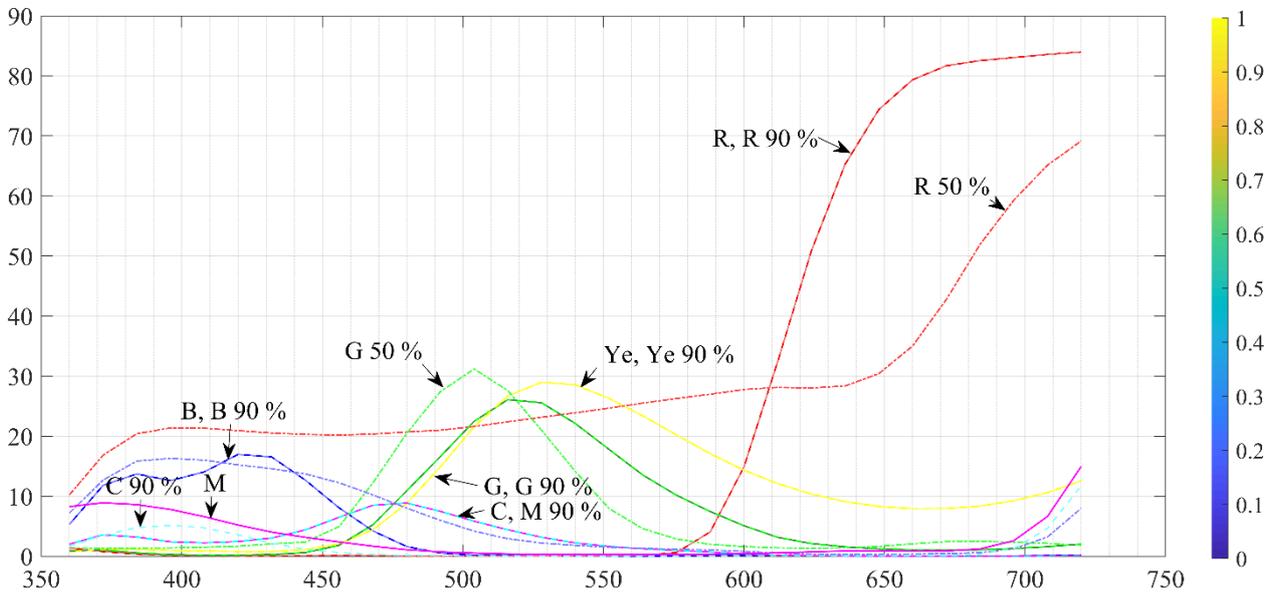

Figure. 5. Spectral distribution of colors depicted in fig. 3

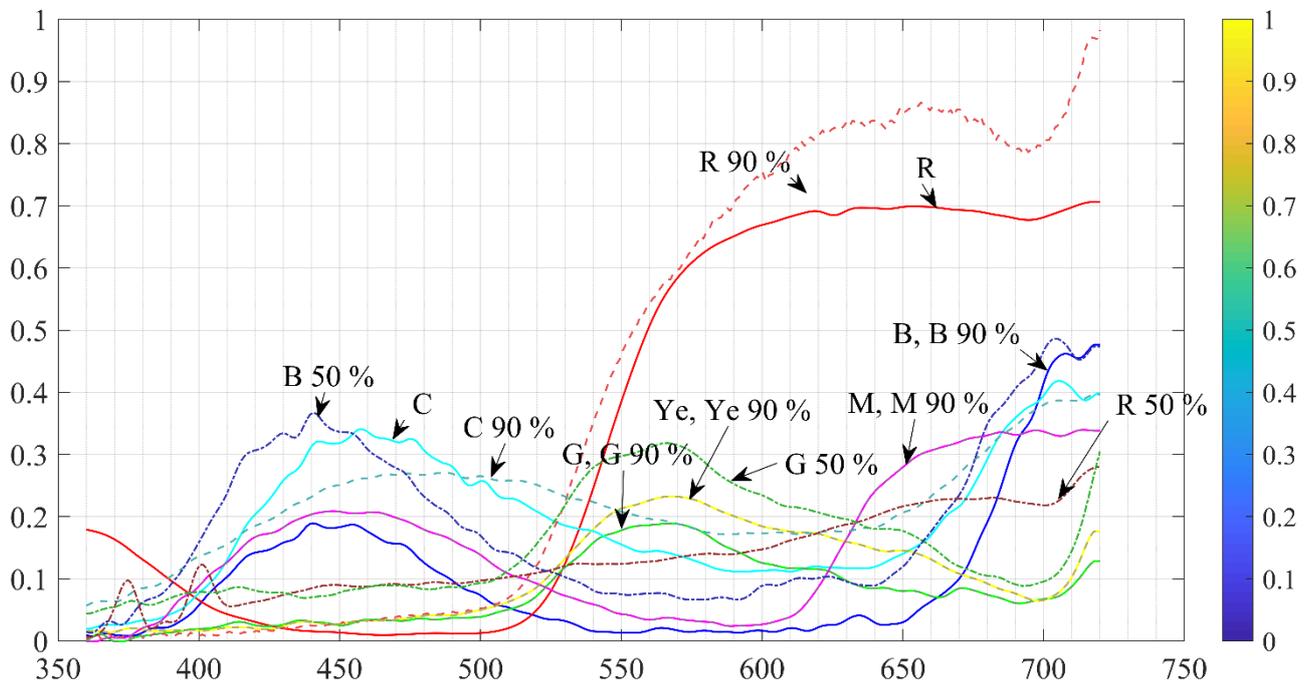

Figure. 6. Spectral distribution of the colors depicted in fig. 4

The spectra are represented in Fig. 5, 6, is the desired result for use in building optical testing tables. Its shortcomings and some, not the matches depicted in Fig. 3, 4, but the existing currently standardized spectral distributions correspond to the chosen colors.

### 4.3. Accounting for dynamic measurement conditions

You should objectively judge how much of the existing color area can be transmitted to TV systems, taking into account the visual perception of color images. To ensure this purpose, it is desirable to represent the region of transmitting colors in the coordinates of the equivalent color space. Below is a boundary of the transmitting color area in the coordinate plane of the equivalent color space, in more detail the model is represented in the works [15, 16]. $a'_M, b'_M$ CAM16$(a'_M, b'_M)$

At the present stage, this posts can be considered the most promising color space for color assessments, based on modern models of color perception. Structural diagram of the model is presented in Fig. 7. This model will be used for further construction of testing color sets.

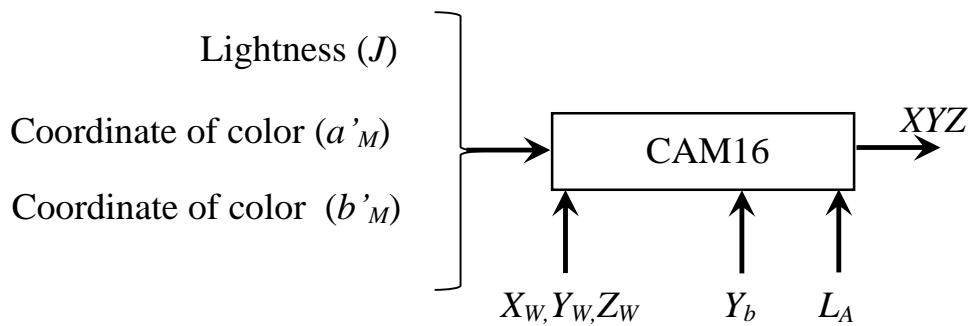

Figure. 7. Input and output parameters of the color appearance model CAM16

In Fig. 7 flagged adapting values: $X_W$, $y_w$, $Z_w$ – colorability coordinates of the lighting source, $y_b$ – color image background brightness, $L_A$– Brightness setting that adjusts eyesight.

Unlike the traditional XYZ coordinate system, the colors in the equivalent coordinate system are at the same distance. This gives an opportunity to argue that the square of the transmitted color system will be able to estimate the specified area. The distance between colors in the work was chosen 2 units IKO, which corresponds to the threshold of Kolororozrìznennâ. This threshold is selected for reasons that if the transmission of color will be with an error, it can be found both objective and subjective method. These color sets are represented in Fig. 8 – 13.

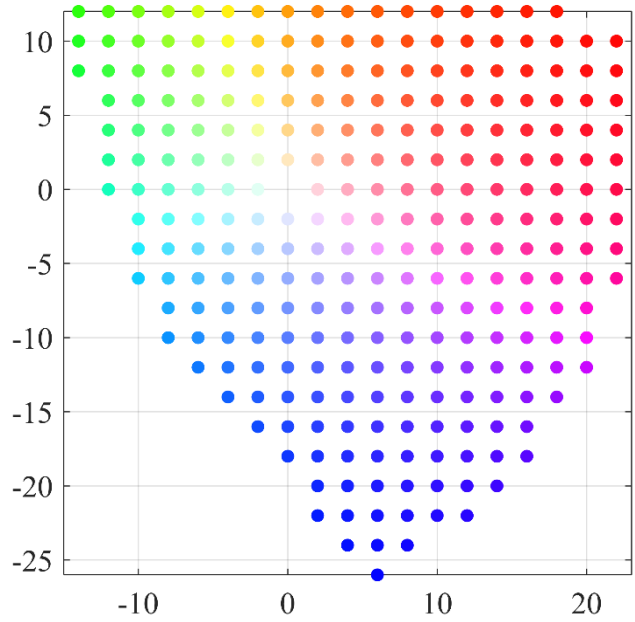

Figure. 8. The color region in the coordinates $a'_m$, $b'_m$, which is transmitted to the TV with the adaptation conditions of $L_A = 50^2$ cd/m², view conditions (VC) – ' Average '
$J = 10$

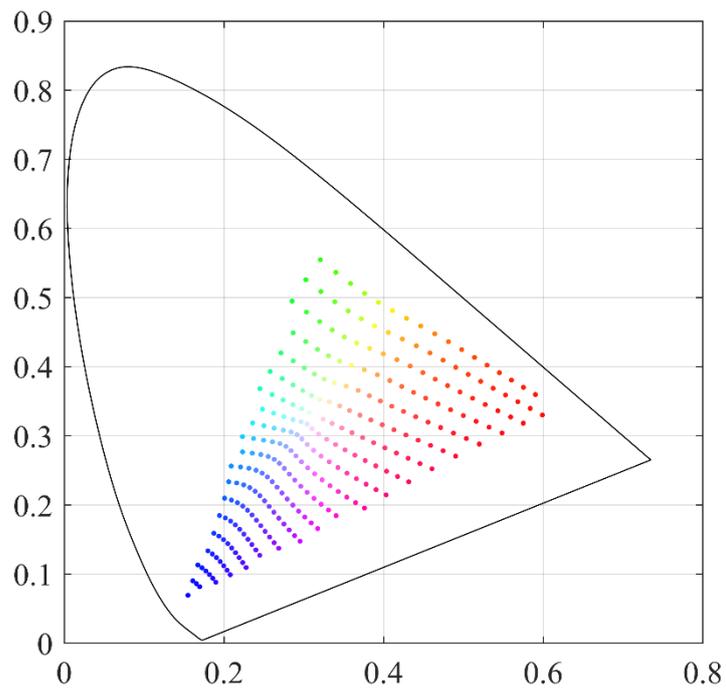

Figure. 9. The color satin points are shown in Fig. 6, presented in a traditional coordinate system *XYZ*

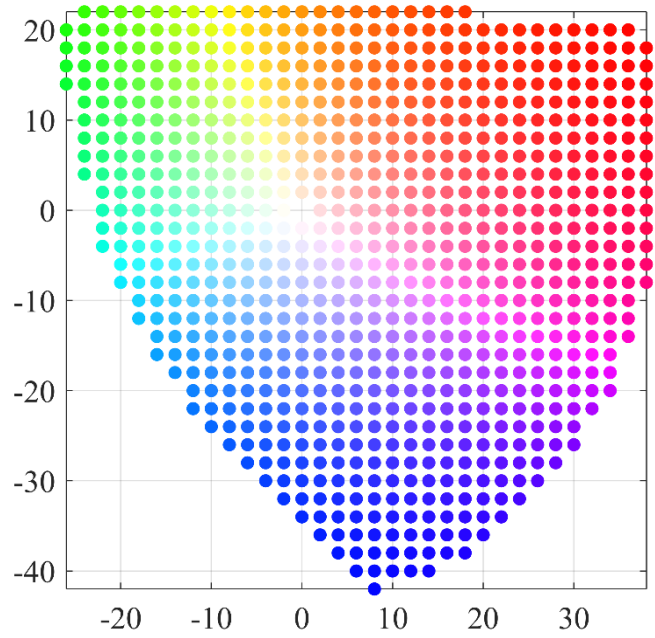

Figure. 10. The color area in the coordinates of $a'_M$, $b'_m$, which is transmitted to a tweeter with the adaptation conditions of $L_A = 50^2$ cd/m², Terms of view (VC) – ' Dark ' $J = 50$

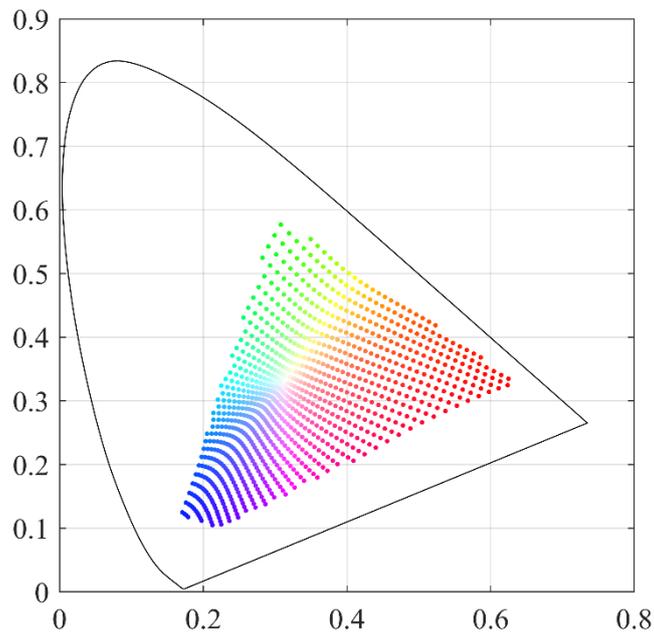

Figure. 11. The color satin dots are shown in Fig. 8, presented in a traditional coordinate system *XYZ*

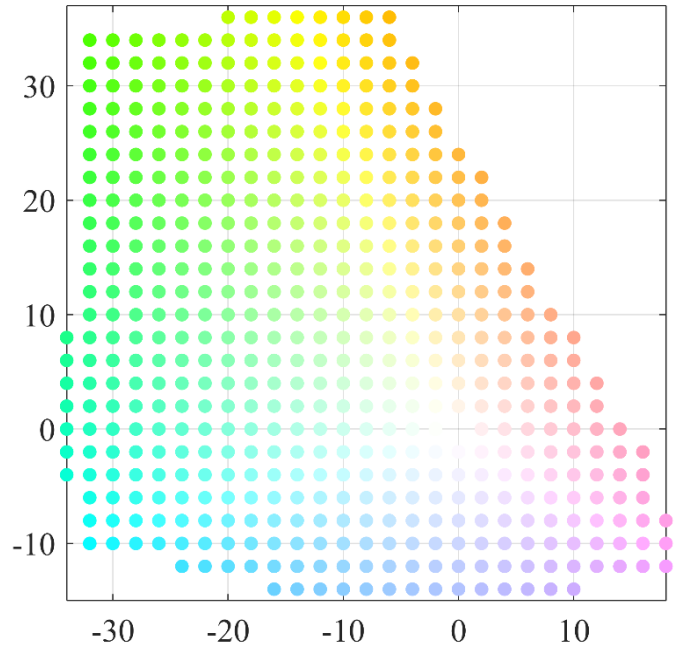

Figure. 12. The color area in the coordinates of $a'_M$, $b'_m$, which is transmitted to a tweeter with the adaptation conditions of $L_A = 50^2$ cd/m², Terms of view (VC) – ' Average ' $J = 90$

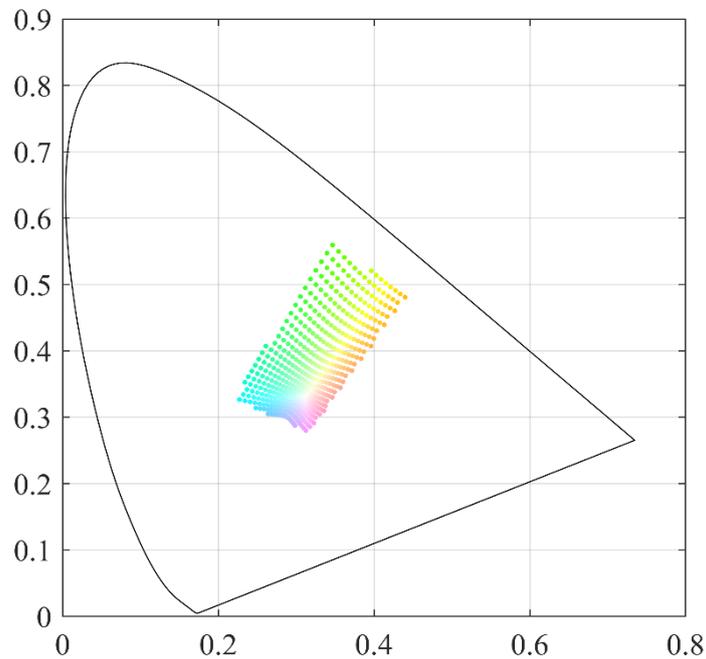

Figure. 13. The color satin points are shown in Fig. 10, presented in a traditional coordinate system *XYZ*

From Fig. 8 – 13 It is noticeable that for different conditions of adaptation and light (*J*) The value of sets of points differs. and notably, in dark and dim terms the number of colors and their threshold distinguishing is much lower than in brighter. These interpretations can be To observe Fig. 6 – 9. While increasing the brightness of the environment, the number of color points increases, but the region of transmitting colors decreases, which is the physical limitation of the image transfer system.

### 5. Discussion of research results proposed adaptation algorithm

According to the optimal value of color dots, it is customary to choose basic and additional colors with relative brightness of 100, 75 and 25%, as shown in P. 4. 2. But their use is insufficient and does not allow to assess the saturated color in full, as well as less saturated colors.

Therefore, the proposed set of color fig. 2, table. 1 is defined on the base of optimum color, which consists of the basic, complementary colors and colors with the saturation of 50%. The proposed algorithm for finding optimal spectral distributions is universal and allows to obtain the spectral distribution of optimal colors for the arbitrary system.

In practice, it is difficult to make optical samples with optimum spectral distribution, which is why it was suggested to define a set of colors from existing sets of spectra. The kits of the proposed spectral distributions are represented in Fig. 3, 4, and the value of their errors in the table. 1. Not exceeding 0.2 CIE units. This set is recommended for evaluation of high definition multimedia systems.

If the task is to determine the most accurate error of transmitted colors, the proposed sets of colors are not enough. Therefore, you need to use satin colors, presented in Fig. 8 – 13. These color atlases are constructed using an equivalent coordinate system which enables to define the necessary set of colors taking into account the characteristics of the human vision.

It should be understood, that in practice can be used test material with a set presented in Fig. 1, 2, or more advanced features. 5, 6, or the most complete fig. 8 – 13. The choice of one or another set depends on the required accuracy of the obtained results and the rate of receipt.

It is also necessary to mention the limitations of this study, in particular, to indicate the boundaries of the application of the proposed solutions both in practical terms and in theoretical.

Further development of these researches can be generators of testing signals, and also for constructing optical testing of tables for estimation of the functioning of multimedia pathts.

### 7. Conclusions

1. It is recommended to use for measurement of colorimetric parameters of digital systems of objective measurement methods. For objective assessments, the development of effective evaluation methods is offered, namely the use of optical testing tables with spectral distributions. For an option of no optical test table, the color atlas coordinates should be presented in the work.

2. According to the results of the study, a set of optimum colors were offered, which at the moment is used to assess their number equal to seven colors, and it is proposed to extend this set to sixteen. The latter set is more effective as it allows to estimate the colors with the saturation 100, 90 and 50%. The color sets and algorithm of their search for the high-definition television system are presented. The error when using the proposed algorithm is $\Delta E \leq 10^{-5}$. At the same time for the spectra of the existing sets, the amount of error is presented not exceeding 0.2 units MKO. For a detailed assessment of colorimetric Properties of the transmission quality and the transmitting color area, use the color set presented in the work.

3. Taken into account the dynamic properties of the adaptive properties of human vision, represented in Fig. 7. Recommendations for the full estimation of colorimetric characteristics should be made using satin colors with distances of 2 units of IKO. This distance is classified by the person as "not noticeable" color changes, so if a discrepancy is noticeable it will give an opportunity to classify the system work, as that distorts the transmitted information. Equal distance between colors achieved using the equivalent CAM16 system.

**Thanks**